# A scalable Controlled NOT gate for linear optical quantum computing using microring resonators

Ryan E. Scott[1], Paul M. Alsing[2], A. Matthew Smith[2], Michael L. Fanto[2], Christopher C. Tison[2], James Schneelolch[2] and Edwin E. Hach, III[1*]

[1]*School of Physics and Astrophysics, Rochester Institute of Technology, 85 Lomb Memorial Drive, Rochester, New York 14623, USA*

[2] *Air Force Research Laboratory, Information Directorate, 525 Brooks Rd., Rome, NY 13411*

*\*corresponding author:eehsps@rit.edu*

Abstract: We propose a scalable version of a KLM CNOT gate based upon integrated waveguide microring resonators (MRR), vs the original KLM-approach using beam splitters (BS). The core element of our CNOT gate is a nonlinear phase-shift gate (NLPSG) using three MRRs, which we examine in detail. We find an expanded parameter space for the NLPSG over that of the conventional version. Whereas in all prior proposals for bulk optical realizations of the NLPSG the optimal operating point is precisely a single zero dimensional manifold within the parameter space of the device, we find conditions for *effective* transmission amplitudes which define a set of one dimensional manifolds in the parameters spaces of the MRRs. This allows for an unprecedented level flexibility in operation of the NLPSG that and allows for the fabrication of tunable MRR-based devices with high precision and low loss.

In 2001, Knill, Laflamme and Milburn (KLM) proposed an efficient scheme for linear optical quantum computing [1]. The KLM proposal is based upon a probabilistic, two-qubit, Controlled NOT (CNOT) gate along with local unitary operations on individual qubits. Some years later, Okamoto, et. al., demonstrated experimentally a realization the KLM CNOT gate in bulk optics [2]. The KLM CNOT gate, shown schematically in Fig. (1), is itself composed of two Non-Linear Phase Shift Gates (NLPSG), the essential two-qubit element of the CNOT gate. Each NLPSG is a probabilistic device involving three optical modes, that, in the bulk optical realization encounter strategically placed and optimally reflective beam splitters that appropriately route the free space evolution of photonic states through the system. The KLM CNOT gate performs a two qubit operation, namely, a flip of the target qubit ($t$) conditioned on the value of the control qubit ($c$), as $|i\rangle_c |j\rangle_t \xrightarrow{\text{CNOT}} |i\rangle_c |i \oplus j\rangle_t$.

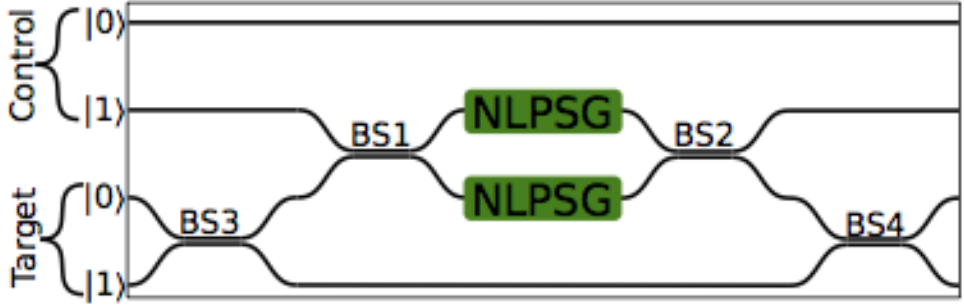

**Figure 1** Schematic diagram of the KLM CNOT gate composed of two NLPSGs.

In the dual rail encoding scheme indicated in Fig. (1), each qubit requires a *single* photon in one of two optical modes; it is a two qubit gate acting on a two photon system. Specifically, the 'bunching' of two photons in any of the individual modes (in or out) is a failure of the gate. That such failures must be rejected is the origin of the probabilistic nature of the gate. The role of each NLPSG is to ensure that states involving two photons in the same mode interfere completely destructively at the next SU(2) (or U(2) as in our proposal here) linear optical element they encounter after being excited in the first place. This is accomplished in the gate shown in Fig. (1) as long as the NLPSG impart a phase shift of $\pi$ radians on the two photon branch of any single mode state that encounters it,

$$|\psi\rangle = \alpha_0 |0\rangle + \alpha_1 |1\rangle + \alpha_2 |2\rangle \xrightarrow{\text{NLPSG}} |\psi'\rangle = \alpha_0 |0\rangle + \alpha_1 |1\rangle - \alpha_2 |2\rangle \qquad (1)$$

wherein normalization of the input, and, therefore, because the coefficients that appear are either the same or shifted by $\pi$ radians, output state requires that $|\alpha_0|^2 + |\alpha_1|^2 + |\pm\alpha_2|^2 = 1$.

There is currently no way known to deterministically effect the transformation in Eq. (1) via unitary evolution. Instead, the transformation is realized probabilistically by using two auxiliary optical modes with one ancilla input photon. Projecting out a specific final state of the two-mode auxiliary subsystem, the nonlinear phase shift produces the desired local isometry on the remaining mode. It has been shown in [1, 3] that this action is successfully with a probability of ¼ and that the result of the projective measurement faithfully indicates the success of the transformation. Consequently, the optimal probability of success for the KLM CNOT gate is $1/16$.

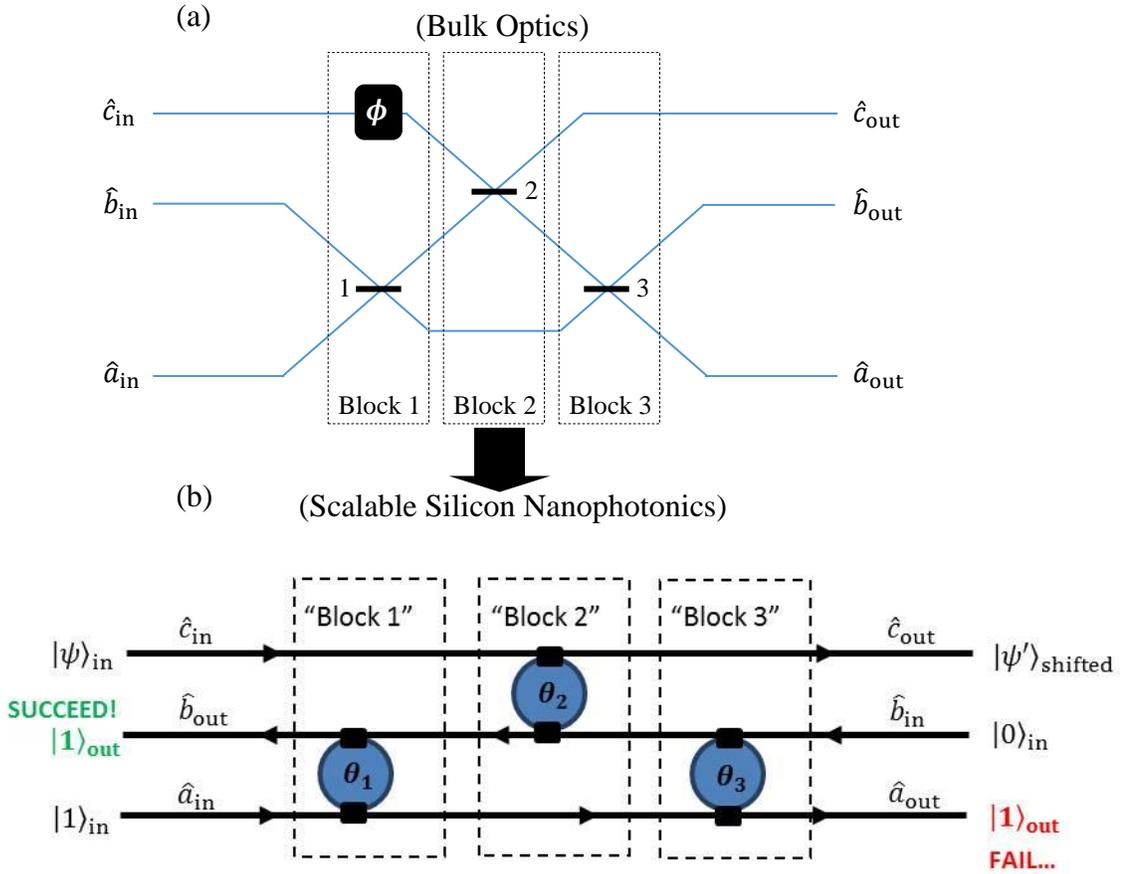

**Figure 2** A schematic diagram of a NonLinear Sign Gate (a) in bulk optics and (b) as implemented via our proposal using directionally coupled silicon nanophotonics waveguides and microring resonators (MRR). The nonlinear sign flip is effected on the state in mode *c*, as given in Eq. (1); modes *a* and *b* are auxiliary modes required for the probabilistic action of the gate. The black arrow connecting parts (a) and (b) of the figure effectively summarizes the advancement we discuss in detail in this paper.

Bulk optical realizations of the KLM CNOT are not scalable, discounting them as potential candidates for components of a viable quantum computer. Separate from scalability, bulk realizations based upon beam splitters and linear phase shifters lack any significant opportunity for dynamical tuning of parameters as might be desirable in a practical operating environment. The ability to 'scan' the parameter space of a device in situ to find a set of parameters allowing for optimal operation, viz. a success probability of ¼ for an NLPSG, allows for further tailoring of device and system design to a specific quantum computation.

Previously, we have predicted the existence of multi-dimensional Hong-Ou-Mandel Manifolds in the operating parameter space of a double bus microring resonator (MRR) [4-7]. This structure, which we identify as a fundamental circuit element for scalable quantum information processing in silicon nanophotonics, admits infinitely many more possibilities for realizing the Hong-Ou-Mandel Effect than does a traditional 50/50 beam splitter (BS) in bulk optics. Further, the double bus ring resonator is inherently scalable and easy to integrate in silicon nanophotonics. In brief, the replacement of each BS by a double bus MRR increases the number the available tunable parameters from one to three (one transmission coefficient for the BS; two transmission coefficients and one round trip phase for the MRR), which greatly expands the overall device parameter space. Thus, as will be demonstrated in this work, the optimal single point solution for the three BS transmission coefficients in the KLM BS version of the NLPSG will be expanded to sets of one and two dimensional manifolds when three MRRs are used.

The purpose of this letter is two-fold; (1) to propose a scalable version of the NLPSG based upon the fundamental circuit element we examined in Ref. [4], as the key nonlinear element of the CNOT gate, and (2) to examine the higher dimensional manifolds within the NLSG parameter space on which the desired nonlinear phase shift occurs with optimal probability ¼. Fig. (2) summarizes our proposal. In Fig. (2a) we show the basic design for an NLPSG in bulk optics; this is essentially the same design as proposed in Ref. [1]. Fig. (2b) shows our scalable version based upon a network of silicon nanophotonic waveguides directionally coupled to microring resonators. The role of the double bus MRR as a circuit element is obvious.

Referring throughout to the labeling scheme introduced in Fig. (2a), the nonlinear phase shift is to occur on the state propagating through the system along the upper rail, $a_{1,\text{in}} \to a_{1,\text{out}}$. The lower two rails support the required auxiliary modes upon which a projective measurement of the output is performed in order to complete the phase shift. The operation of the NLPSG proceeds as follows. The system input is prepared in the global state,

$$|\Psi\rangle = |\psi\rangle_1 \otimes |1\rangle_2 \otimes |0\rangle_3 \qquad (2)$$

The global output state resulting from purely unitary evolution *can* be written in the form (see Supplemental Material further details),

$$|\Psi'\rangle = \hat{U}|\Psi\rangle = \beta|\Psi^{NLPSG}\rangle + \sqrt{1-|\beta|^2}\,|\Psi^\perp\rangle \qquad (3)$$

where $0 \le |\beta| \le 1$, $\hat{U}$ describes the unitary evolution of the global system from input→output, $|\Psi^{NLPS}\rangle$ is the branch of the output state that induces the nonlinear phase shift upon projective measurement, and $|\Psi^\perp\rangle$ is the branch that is rejected by the measurement such that $\langle \Psi^\perp | \Psi^{NLPS}\rangle = 0$. The probability of success for the NLPSG is given by,

$$p_{\text{success}}^{NLPS} = \langle \Psi' | \hat{P}^{NLPSG} | \Psi' \rangle = |\beta|^2. \qquad (4)$$

Using linear optical transformations as in [3],

$$\hat{a}_{j,\text{in}}^\dagger \to \hat{U}\hat{a}_{j,\text{in}}^\dagger \hat{U}^\dagger = \sum_{k=1}^{3} S^T_{jk}\hat{a}_{k,\text{out}}^\dagger = \sum_{k=1}^{3} S_{kj}\hat{a}_{k,\text{out}}^\dagger \qquad (5)$$

where the coefficients $S^T_{jk}$ encode the scattering of of the input→output modes which depend on the system parameters labeled in Fig. (2b). We arrive at the following set of constraints that result in the successful implementation of the NLPSG,

$$\alpha_0 S_{22} = \beta \alpha_0, \tag{6}$$

$$\alpha_1 \left( S_{11} S_{22} + S_{21} S_{12} \right) = \beta \alpha_1, \tag{7}$$

$$\alpha_2 S_{11} \left( S_{11} S_{22} + 2 S_{21} S_{12} \right) = -\beta \alpha_2. \tag{8}$$

Mathematical consistency between Eqs.(7) and (8) requires that $S_{11} = 1 \pm \sqrt{2}$, physical consistency, $|S_{11}| \leq 1$, further restricts this to the value,

$$S_{11} = 1 - \sqrt{2}. \tag{9}$$

Combining Eqs. (6) - (9),

$$p_{\text{success}}^{NLPSG} = |\beta|^2 = |S_{22}|^2 = \tfrac{1}{2} |S_{21}|^2 |S_{12}|^2. \tag{10}$$

For an optimal choice of linear optical couplings, namely *one* particular combination of beam splitter reflectivities in bulk optical implementations, see Fig. (2a), it has long been established that the maximum possible probability of success for the NLPSG is $p_{\text{success,max}}^{NLPS} = 1/4$ [1,3]

In order to analyze the operation of the scalable NLPSG shown in Fig. (2a), we must work out the coefficients $S_{ij}$ in order to get $S_{ij}^T = S_{ji}$, that describes the linear transformation of creation operators which characterizes the unitary evolution of the three-mode fields through the device. Previously, we have proposed the directionally coupled double bus MRR as a fundamental circuit element for optical quantum information processing in silicon nanophotonics [4]. Comparing Fig. (2a) with Fig. (1) from Ref. [4], it is clear that our proposed NLPSG is an integrated network of three such fundamental circuit elements. To aid in the discussion we group the NLPSG into three "blocks," with each block involving a single MRR based circuit element coupling two of the modes; the remaining mode involves a linear phase shift across any given block. For example, regarding modes 2 and 3 of block 1 in Fig. (2a), the input Boson operators, $\hat{a}_{3,\text{in}}$ and $\hat{b}_{21}$, to the MRR are analogous to the operators $\hat{a}$ and $\hat{f}$, respectively, in Ref. [4], with a similar analogy for the outputs $\hat{a}_{12}, \hat{a}_{2,\text{out}} \leftrightarrow \hat{c}, \hat{l}$. Along mode 1, $\hat{c}_{12} = e^{-i\delta_1} \hat{a}_{1,\text{in}}$. Boson operators carrying two subscripts are internal to the NLPSG; we will eliminate them algebraically in deriving the operator input/output relations for the device.

In the notation we have adopted here the input/output operator transformations for the individual fundamental circuit elements implicated in Fig. (2a) can be written as,

$$\begin{pmatrix} \hat{a}_{2,\text{out}} \\ \hat{a}_{12} \end{pmatrix} = \mathbf{M}^{(1)} \begin{pmatrix} \hat{b}_{21} \\ \hat{a}_{1,\text{in}} \end{pmatrix}, \quad \begin{pmatrix} \hat{c}_{23} \\ \hat{b}_{21} \end{pmatrix} = \mathbf{M}^{(2)} \begin{pmatrix} \hat{c}_{12} \\ \hat{b}_{32} \end{pmatrix}, \quad \begin{pmatrix} \hat{b}_{32} \\ \hat{a}_{3,\text{out}} \end{pmatrix} = \mathbf{M}^{(3)} \begin{pmatrix} \hat{a}_{2,\text{in}} \\ \hat{a}_{23} \end{pmatrix} \tag{11}$$

Where the superscript on the $2 \times 2$ matrix $\mathbf{M}^{(j)}$ labels the MRR with which it corresponds. Each of these matrices has the form,

$$\mathbf{M}^{(j)} = \begin{pmatrix} A_j & B_j \\ C_j & D_j \end{pmatrix} \tag{12}$$

having matrix elements [4],

$$A_j = \frac{\eta_j - \tau_j^* e^{-i\theta_j}}{1 - \eta_j^* \tau_j^* e^{-i\theta_j}}, \qquad B_j = -\frac{\gamma_j \kappa_j^* e^{-i\phi_j}}{1 - \eta_j^* \tau_j^* e^{-i\theta_j}} \tag{13}$$

$$C_j = -\frac{\kappa_j \gamma_j^* e^{-i(\theta_j - \phi_j)}}{1 - \eta_j^* \tau_j^* e^{-i\theta_j}}, \qquad D_j = \frac{\tau_j - \eta_j^* e^{-i\theta_j}}{1 - \eta_j^* \tau_j^* e^{-i\theta_j}} \tag{14}$$

that depend explicitly on the system parameters, namely, the round trip phase shifts $(\theta_j)$ of a ring, the direct transmission amplitudes $(\tau_j, \eta_j)$, and the cross coupling amplitudes $(\kappa_j, \gamma_j)$ at the directional coupler. The other phase shifts $\phi_j$ represents the phase partition induced by the specific locations of the couplings of the rings with the waveguides; the phase partitions have no effect on our results, so we implicitly set them to $\phi_j = \theta_j/2$ (symmetrically coupled rings).

There is no direct algebraic substitution that will result in an operator input/output relation of the form we desire, namely,

$$\begin{pmatrix} \hat{a}_{1,\text{out}} \\ \hat{a}_{2,\text{out}} \\ \hat{a}_{3,\text{out}} \end{pmatrix} = \mathbf{S} \begin{pmatrix} \hat{a}_{1,\text{in}} \\ \hat{a}_{2,\text{in}} \\ \hat{a}_{3,\text{in}} \end{pmatrix} \tag{15}$$

where the matrix $\mathbf{S}$ describes the unitary "scattering" of the input operators into the output ones. Instead, owing to the directional nature of the couplings between waveguides and MRRs and to the topology of each MRR itself, we must algebraically adjust the relations encoded in transfer matrices (described in detail in the Supplemental Material) by introducing a set of three mode swap operations. The Bosonic commutation relations $\left[\hat{a}_{j,\text{out}}, \hat{a}_{k,\text{out}}^\dagger\right] = \delta_{jk}$, $\left[\hat{a}_{j,\text{out}}, \hat{a}_{k,\text{out}}\right] = \left[\hat{a}_{j,\text{out}}^\dagger, \hat{a}_{k,\text{out}}^\dagger\right] = 0$ with similar relations for the input operators, constrain the $\mathbf{S}$ matrix to be unitary, $\mathbf{S}^{-1} = \mathbf{S}^\dagger = \left(\mathbf{S}^T\right)^*$, which, in turn, implies that $\left(\mathbf{S}^{-1}\right)^* = \mathbf{S}^T$ so that we arrive at a description of the input creation operators for the MRR in terms of the output creation operators

$$\begin{pmatrix} \hat{a}_{1,\text{in}}^\dagger \\ \hat{a}_{2,\text{in}}^\dagger \\ \hat{a}_{3,\text{in}}^\dagger \end{pmatrix} = \mathbf{S}^T \begin{pmatrix} \hat{a}_{1,\text{out}}^\dagger \\ \hat{a}_{2,\text{out}}^\dagger \\ \hat{a}_{3,\text{out}}^\dagger \end{pmatrix} \tag{16}$$

analogous to Eq. (5).

In order to preserve Bosonic commutation relations, each of the directional couplers must obey the reciprocity relations [8],

$$\begin{aligned} |\kappa_j|^2 + |\tau_j|^2 &= 1 \\ \kappa_j \tau_j^* + \kappa_j^* \tau_j &= 0 \end{aligned} \quad (17)$$

with similar relations for $\kappa_j \to \gamma_j$ and $\tau_j \to \eta_j$. The engineering of directional couplers is such that the direct transmission amplitudes are real $(\tau_j, \eta_j) \to (|\tau_j|, |\eta_j|)$. According to Eqs. (17) then,

$$\begin{aligned} \kappa_j &= i\sqrt{1-|\tau_j|^2} \\ \gamma_j &= i\sqrt{1-|\eta_j|^2} \end{aligned}. \quad (18)$$

Regarding the device we propose in Fig. (2a), this apparently leaves three (3) round trip phase shifts, $\theta_j$, three (3) in line phase shifts, $\delta_j$, and six (6) coupling parameters, $(|\tau_j|, |\eta_j|)$ for a total of twelve (12) physical design parameters to describing the system. Two of the in line phase shifts, namely, $\delta_1$ and $\delta_3$, seem to be external and, therefore, superfluous, but, as we shall discuss below, are actually required to tune the system in certain ways to effectively compensate for the internal phase shift $\delta_2$, which is in no way superfluous. Nevertheless, owing to the compensatory role they play, we shall omit $\delta_1$ and $\delta_3$ from the list of design and optimization parameters we consider, paring the set of these down to ten (10).

In light of Eq. (9), a little bit of simple algebra reveals that Eqs. (7) and (8) are identical constraints; in other words these two equations place two constraints, one on the real parts and the other on the imaginary parts, on the complex elements of **S** matrix. Similarly, Eq. (6) itself places two (2) more independent constraints on the system for a total of four (4) constraints due to the specific required action of the NLPSG. Interjecting all of this into our accounting from the previous paragraph, we arrive at a total of six (6) free design and optimization parameters for our proposed NLPSG. In a similar vein to our already published results regarding the Hong-Ou-Mandel Effect [2014], we now search for $N$ dimensional manifolds within the parameter space of the device, where $N \leq 6$, upon which the NLPSG operates with the theoretically maximal probability of success of $\frac{1}{4}$.

We shall seek solutions with maximum success probability mimicking the solution and procedure of the bulk beam splitter NLPSG (see [3]) as if each MRR in Fig.(2b) were collapsed to a BS as in Fig. (2a) with an effective transmission and reflection coefficients *t, r* respectively. We will first treat analytically the optimal operating conditions on a one dimensional manifold for which the MRRs are all set to be on resonance, $\theta_i = 2\pi$, with balanced phase partitioning, $\phi_i = \theta_i/2 = \pi$, and all linear phase shifts $\delta_i = 0$ for $i = \{1, 2, 3\}$. Under these conditions, Eqs. (13) and (14) take the forms,

$$\mathbf{M}^{(1)} = \begin{pmatrix} t_1 & r_1 \\ r_1 & -t_1 \end{pmatrix}, \mathbf{M}^{(2)} = \begin{pmatrix} -t_2 & r_2 \\ r_2 & t_2 \end{pmatrix}, \text{ and } \mathbf{M}^{(3)} = \begin{pmatrix} t_3 & r_3 \\ r_3 & -t_3 \end{pmatrix} \tag{19}$$

where we have introduced for each MRR real, effective transmission and reflection coefficients, $t_i$ and $r_i$, respectively with,

$$t_i \equiv \frac{\eta_i - \tau_i}{1 - \eta_i \tau_i} \tag{20}$$

and, referring to Eqs. (17) and (18),

$$r_i = \frac{\sqrt{(1-|\tau_i|^2)(1-|\eta_i|^2)}}{1 - \eta_i \tau_i}, \tag{21}$$

from which is it straightforward to show that $r_i^2 + t_i^2 = 1$. Note that for each MRR, $-1 \leq \eta_i, \tau_i \leq 1$ are the *physical* upper and lower transmission coefficients, while $-1 \leq t_i \leq 1$ is a parameter that has the form of an effective transmission coefficient. Equation (20) then defines a 1D manifold $\eta_i(\tau_i; t_i) \equiv (t_i + \tau_i)/(1 + t_i \tau_i)$ parameterized by $t_i$. The **S** matrix relating the input Boson operators to the ouputs as desired for the Heisenberg Picture description of the unitary part of the evolution of the NLPSG via Eq. (16) then takes the form, after some lengthy algebra,

$$\mathbf{S} = \frac{1}{t_2\sqrt{1-t_1^2}\sqrt{1-t_3^2} - 1} \begin{pmatrix} t_2 - \sqrt{1-t_1^2}\sqrt{1-t_3^2} & -t_3\sqrt{1-t_2^2} & t_1\sqrt{1-t_2^2}\sqrt{1-t_3^2} \\ -t_1\sqrt{1-t_2^2} & -t_1 t_2 t_3 & t_2\sqrt{1-t_3^2} - \sqrt{1-t_1^2} \\ t_3\sqrt{1-t_1^2}\sqrt{1-t_2^2} & t_2\sqrt{1-t_1^2} - \sqrt{1-t_3^2} & -t_1 t_3 \end{pmatrix} \tag{22}$$

such that $\det(\mathbf{S}) = -1$.

In order to analyze the operations of the NLPSG under the foregoing conditions, we must apply the constraints that induce the desired local isometry on the unitarily evolved state in the target mode, Mode 1. Specifically, Eq. (9) requires that $S_{11} = 1 - \sqrt{2}$, which, combined with Eq.(7) (or (8)) and recalling that $\beta$ is real in this case, implies that,

$$\beta = \frac{S_{12} S_{21}}{\sqrt{2}}. \tag{23}$$

Clearly, Eq. (6) further requires that,

$$\beta = S_{22}. \tag{24}$$

Using the explicit forms of the matrix elements given in Eq.(22) we can find conditions on the *effective* transmission amplitudes, $t_i$, for that satisfy the constraints given by Eqs. (9), (23), and (24). Specifically, we find a fixed solution,

$$t_2 = \frac{1+2\sqrt{2}}{7} \approx 0.546918 \tag{25}$$

along with the relationships,

$$t_3 = \left[\frac{\left(2\sqrt{3\sqrt{2}-4}-t_1\right)\left(2\sqrt{3\sqrt{2}-4}+t_1\right)}{(1-t_1)(1+t_1)}\right]^{\frac{1}{2}} \tag{26}$$

and

$$|\beta|^2 = \frac{(1+\sqrt{2})(12\sqrt{2}-16-t_1^2)t_1^2}{16(1-t_1)(1+t_1)}. \tag{27}$$

Optimizing $|\beta|^2$ with respect to $t_1$ by solving $\frac{\partial}{\partial t_1}|\beta|^2\Big|_{t_1=\mathcal{T}_1} = 0$ yields $|\beta|^2_{\max} = 1/4$ for the optimal value of $t_1 \to t_{1,\text{optimal}} \equiv \mathcal{T}_1 = \sqrt{2(\sqrt{2}-1)} \approx 0.91018$. Recalling that the probability of success for the NLPSG is $|\beta|^2$, the maximum value we obtain here agrees completely with that originally posed by Knill Laflamme and Milburn. Further, using the optimal value, $\mathcal{T}_1$, in Eq. (26), we find the optimal value for the remaining effective transmission amplitude, $t_3 \xrightarrow{\text{optimize}} \mathcal{T}_3 = \mathcal{T}_1 = \sqrt{2(\sqrt{2}-1)} \approx 0.91018$.

Summarizing what we have found so far, under conditions of exact resonance and balanced phase partitioning of the MRRs, in-line phase shifts of $0 \bmod 2\pi$ along all waveguides, and real direct transmission amplitudes at all directional couples, the circuit shown in Fig. (2b) will successfully perform a nonlinear sign flip on Mode 1 with a maximum possible probability of success of $1/4$ whenever the effective transition amplitudes for the MRRs are tuned to the optimal values,

$$\begin{aligned}\mathcal{T}_1 = \mathcal{T}_3 &= \sqrt{2(\sqrt{2}-1)} \\ \mathcal{T}_2 &= \frac{1+2\sqrt{2}}{7}\end{aligned}. \tag{28}$$

All of this is in direct correspondence with the results of KLM and Skaar regarding the optimal operating *point* for a NLPSG.

Here is the central point of our work. Whereas, in all prior proposals or realizations of the NLPSG the optimal operating point is precisely that, a single zero dimensional manifold within the parameter space of the device, the conditions placed by Eq.(28) on the *effective* transmission amplitudes define curves, i.e. one dimensional manifolds, in the parameters spaces of the MRRs. This allows for an unprecedented level flexibility in operation of the NLPSG that we propose.

To see how this arises, recall Eq. (20) defining the effective transmission amplitudes for the MRRs. Substituting the fixed optimal values from Eq. (28) for the effective transmission amplitudes results in optimal operating curves for each of the MRRs,

$$\mathcal{T}_i = \frac{\eta_i - \tau_i}{1 - \eta_i \tau_i} \Rightarrow \eta_i(\tau_i; \mathcal{T}_i) = \frac{\mathcal{T}_i + \tau_i}{1 + \mathcal{T}_i \tau_i}. \tag{29}$$

Eq. (29) is effectively the engineering blue print for the optimal operation of the scalable NLPSG we propose.

In Fig. (3) we plot the one dimensional manifolds $\eta_i^2(\tau_i; T_i)$ vs $\tau_i^2$ ($i = \{1,3\}$, black solid; $i = 2$, black dashed) obtained from Eq. (29), for optimal operation of the scalable NLPSG as determined via the conditions developed in Eqs. (23) through (28). It is important to note that even though the effective transmission coefficients for the outer MRRs are equal $\mathcal{T}_1 = \mathcal{T}_3$, this does not imply that the physical transmission coefficients are necessarily equal, since $\tau_1 \neq \tau_3 \Rightarrow \eta_1 \neq \eta_3$. Thus, for each MRR there exists the freedom to choose $\tau_i$ independently. This is in stark contrast to the *single* fixed point solution $(\tau_1, \tau_2, \tau_3)$ in the case of the bulk optics KLM BS-based version of the NLPSG. In addition to the inherent scalability of an MRR-based KLM NLPSG, this result emphasizes the dynamic tunability that arises due to the expanded available parameter space for the device.

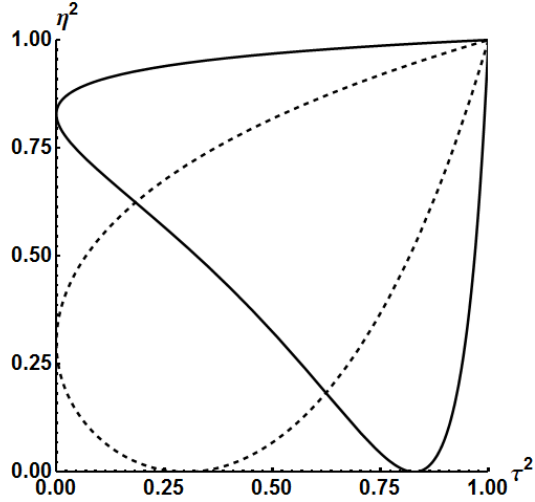

**Figure 3**: The one dimensional manifolds $\eta_i^2(\tau_i; T_i)$ vs $\tau_i^2$ ($i = \{1,3\}$, black solid; $i = 2$, black dashed) obtained from Eq. (29) on which optimal operation ($|\beta|^2 = 1/4$) of the scalable NLPSG occurs under conditions of resonant ($\theta_i = 0 \mod 2\pi$), balanced MRRs and $\delta_i = 0 \mod 2\pi$ phase shifts in the waveguides. In contrast with bulk optical realizations, these curves provide theoretical evidence for vastly enhanced flexibility in implementation and integration of the NLPSG based on directionally coupled MRRs in silicon nanophotonics.

We next try to find other optimal solutions ($|\beta|^2_{\max} = 1/4$) about the above *on-resonance* ($\theta_i = 0 \mod 2\pi$) solutions. To this end, we set $t_i = \mathcal{T}_i$ and $\delta_i = 0$, but allow the MRRs to be off-resonance ($\theta_i \neq 0 \mod 2\pi$). The latter condition implies that the coefficients $A_i, B_i, C_i, D_i$ of Eqs. (13) and (14) are complex. A detailed analysis [9] shows that this yields $\exp(i\theta_{A_2}) = 1$ and $\beta = 1/2 \exp[-i(\theta_{A_1} + \theta_{A_3})]$ (where we have written $A_i = |A_i|\exp[i\theta_{A_i}]$) which translates into (i) $\theta_2 = 0 \mod 2\pi$, but with (ii) $\theta_{A_1}(\eta_1, \tau_1, \theta_1)$, $\theta_{A_3}(\eta_3, \tau_3, \theta_3)$ arbitrary. Condition (i) leads to the same curve as in Fig. 3 for $\tau_2$, $\eta_2(\tau_2; \mathcal{T}_2)$. Condition (ii) implies that arbitrary choices of $\eta_{1,3}, \tau_{1,3}, \theta_{1,3}$ lead to arbitrary values of the phase $\theta_{A_{1,3}}$, and hence these variables are simply constrained by their amplitudes $(\mathcal{T}_{i=1,3})^2 = |A_i(\eta_i, \tau_i, \theta_i)|^2$. These latter equations implicitly define two dimensional surfaces,

$$(\mathcal{T}_{1,3})^2 = \frac{|\eta_{1,3}|^2 + |\tau_{1,3}|^2 - 2|\eta_{1,3}||\tau_{1,3}|\cos\theta_{1,3}}{1 + |\eta_{1,3}|^2|\tau_{1,3}|^2 - 2|\eta_{1,3}||\tau_{1,3}|\cos\theta_{1,3}} \tag{30}$$

as shown in Fig. 4. Note that the center cross section $\theta_{1,3} = 0$ is squared version of Eq.(48) for $t_{1,3} = \mathcal{T}_{1,3}$ which reproduce the $\tau_{1,3}, \eta_{1,3}(\tau_{1,3}; \mathcal{T}_{1,3})$ curves in Fig. 3.

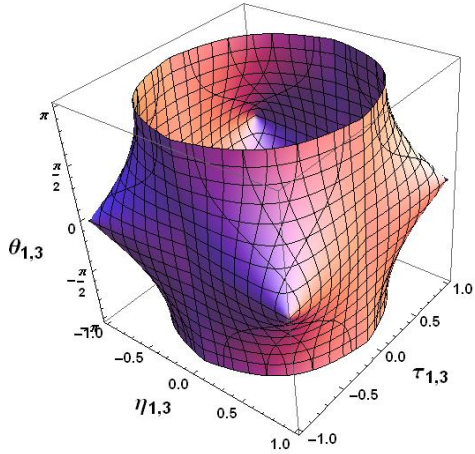

**Figure 4**: The two dimensional manifolds from Eq. (30) on which optimal operation ($|\beta|^2 = 1/4$) of the scalable NLPSG occurs under conditions of on-resonance $\theta_2 = 0 \mod 2\pi$ in the middle MRR, but off-resonance $\theta_{1,3} \neq 0 \mod 2\pi$ in the outer two MRRs (again with $\delta_i = 0 \mod 2\pi$ phase shifts in the waveguides).

In the Supplemental Material we explore the role of the non-trivial inter-MRR phase shift $\delta_2 \neq 0 \mod 2\pi$ for the NLPSG while keeping all the MRRs on-resonance ($\theta_i = 0 \mod 2\pi$, $\delta_1 = \delta_3 = 0$) and again using $t_i = \mathcal{T}_i$ to ensure that $|\beta|^2 = 1/4$. Note that in addition to the freedom to choose $\tau_1, \tau_3,$ and $\eta_1, \eta_3$ (as in Eq. (29)), one now has the freedom to also chose $\theta_1, \theta_3$ independently such that Eq.(30) is satisfied. We note that for the CNOT gate constructed from two NLPSG gates the outer-leg MRR phases $\delta_1, \delta_3$ becomes physically relevant.

The net result of the (non-exhaustive) solutions presented in Figs.(3,4) (and Fig. S1 in the Supplemental Material) is that by using MRRs (as in Fig. 2b) instead of BSs one can achieve optimal NLPSG operation ($|\beta|^2 = 1/4$) while retaining great flexibility in the choice of the parameters $\eta_i, \tau_i, \theta_i$ for each individual MRR, determining its (upper and low) coupling transmission coefficients and phase delay (nearness to resonance). This is to be contrasted with the single-point solution for the triple of BS transmission coefficients in the conventional NLPSG configuration in Fig. 2a. In addition, current silicon (and SiN) CMOS foundry technology (e.g. American Institute of Manufacturing) allows for the fabrication of tunable MRR-based devices with high precision and low loss. The analysis shown here for the increased dimensionality for operational parameter space of the NLPSG as the core building block for an integrated waveguide MRR-based CNOT gate, makes this a promising avenue for other quantum information processing devices utilizing MRR.

**Acknowledgements**

RES and EEH would like to acknowledge support for this work was provided by the Air Force Research Laboratory (AFRL) Visiting Faculty Research Program VFRP GI Grant No. FA8750-16-3-6003. PMA, AMS, MLF, CCT and JS would like to acknowledge support of this work from Office of the Secretary of Defense (OSD) Applied Research for Advanced Science and Technology (ARAP) Quantum Science and Engineering Program (QSEP) program. JS would like to acknowledge support for the National Research Council (NRC). Any opinions, findings and conclusions or recommendations expressed in this material are those of the author(s) and do not necessarily reflect the views of Air Force Research Laboratory.

**References**

[1] E. Knill, R. Laflamme, & G. J. Milburn, *Nature* **409**, 46 (2001).
[2] R. Okamoto, J. L. O'Brien, H. F. Holger, & S. Takeuchi, www.pnas.org/cgi/doi/10.1073/pnas.1018839108, (2011).
[3] J. Skaar, J. C. G. Escartín, and H. Landro, *Am. J. Phys*. **72**, 1385 (2004).
[4] E. E. Hach III, S. F. Preble, A.W. Elshaari, P. M. Alsing, and M. L. Fanto, *Phys. Rev*. **A89**, 043805 (2014).
[5] P. M. Alsing, E. E. Hach III, C. C. Tison, and A. M. Smith, *Phys. Rev.* A **95**, 053828 (2017).
[6] P. M. Alsing and E. E. Hach III, *Phys. Rev.* A**96**, 033847 (2017).
[7] P. M. Alsing and E. E. Hach III, *Phys. Rev.* A**96**, 033848 (2017).
[8] C. C. Gerry and P. L. Knight, *Introductory Quantum Optics* (Cambridge, 2005).
[9] P. M. Alsing and E. E. Hach III, DOI: 10.1117/12.2325080, Quantum Technologies and Quantum Information Science **IV**, Berlin (Sep, 2018).

# Supplemental Material for "A scalable Controlled NOT gate for linear optical quantum computing using microring resonators"

Ryan E. Scott[1], Paul M. Alsing[2], A. Matthew Smith[2], Michael L. Fanto[2], Christopher C. Tison[2], James Schneelolch[2] and Edwin E. Hach, III[1*]

[1]School of Physics and Astrophysics, Rochester Institute of Technology, 85 Lomb Memorial Drive, Rochester, New York 14623, USA

[2] Air Force Research Laboratory, Information Directorate, 525 Brooks Rd., Rome, NY 13411

*corresponding author:eehsps@rit.edu

*Derivation of output state of the NLPSG*

Referring throughout to the labeling scheme introduced in Fig. (2a), the nonlinear phase shift is to occur on the state propagating through the system along the upper rail, $a_{1,\text{in}} \to a_{1,\text{out}}$. The lower two rails support the required auxiliary modes upon which a projective measurement of the output is performed in order to complete the phase shift. The operation of the NLPSG proceeds as follows. The system input is prepared in the global state

$$|\Psi\rangle = |\psi\rangle_1 \otimes |1\rangle_2 \otimes |0\rangle_3 \qquad (31)$$

The global output state resulting from purely unitary evolution *can* be written in the form

$$|\Psi'\rangle = \hat{U}|\Psi\rangle = \beta |\Psi^{NLPSG}\rangle + \sqrt{1-|\beta|^2}\,|\Psi^{\perp}\rangle \qquad (32)$$

where $0 \leq |\beta| \leq 1$, $\hat{U}$ describes the unitary evolution of the global system from input→output, $|\Psi^{NLPS}\rangle$ is the branch of the output state induces the nonlinear phase shift upon projective measurement, and $|\Psi^{\perp}\rangle$ is the branch that is rejected by the measurement such that $\langle \Psi^{\perp}|\Psi^{NLPS}\rangle = 0$. The projection operator

$$\hat{P}^{NLPSG} \equiv \hat{1}^{(c)} \otimes \hat{P}^{(2,3)}_{1,0} \qquad (33)$$

where

$$\hat{P}^{(2,3)}_{1,0} \equiv \left[|1\rangle_2 \otimes |0\rangle_3\right]\left[{}_2\langle 1| \otimes {}_3\langle 0|\right] \qquad (34)$$

Characterizes the successful operation of the NLPSG. Just prior to the measurement, the state of the system can be written as

$$|\Psi'\rangle = \hat{U}|\Psi\rangle = \hat{P}^{NLPSG}\hat{U}|\Psi\rangle + \left(\hat{I} - \hat{P}^{NLPSG}\right)\hat{U}|\Psi\rangle \qquad (35)$$

The state of the system after a measurement in which a single photon is detected in output mode 2 and no photons are detected in output mode 3, is given by

$$\left|\Psi^{NLPSG}\right\rangle = \frac{\hat{P}^{NLPSG}\left|\Psi'\right\rangle}{\sqrt{\left\langle\Psi'\right|\hat{P}^{NLPSG}\left|\Psi'\right\rangle}}; \tag{36}$$

the probability of success for the NLPSG is given by

$$p_{\text{success}}^{NLPSG} = \left\langle\Psi'\right|\hat{P}^{NLPSG}\left|\Psi'\right\rangle = \left|\beta\right|^2 \tag{37}$$

Comparing Eq. (32) with the defining equation for the NLPSG (Eq. (1) in the main text) and (31), successful operation of the NLPSG requires that

$$\left|\Psi^{NLPSG}\right\rangle = \left|\psi'\right\rangle_1 \otimes \left|1\right\rangle_2 \otimes \left|0\right\rangle_3 = \left(\alpha_0\left|0\right\rangle_1 + \alpha_1\left|1\right\rangle_1 - \alpha_2\left|2\right\rangle_1\right) \otimes \left|1\right\rangle_2 \otimes \left|0\right\rangle_3 \tag{38}$$

Unitary evolution of the global state vector produces

$$\left|\Psi'\right\rangle = \hat{U}\left|\psi\right\rangle_1 \otimes \left|1\right\rangle_2 \otimes \left|0\right\rangle_3 = \hat{U}\left(\alpha_0\left|0\right\rangle_1 + \alpha_1\left|1\right\rangle_1 + \alpha_2\left|2\right\rangle_1\right) \otimes \left|1\right\rangle_2 \otimes \left|0\right\rangle_3 \tag{39}$$

$$\left|\Psi'\right\rangle = \left(\alpha_0 + \alpha_1\left\{\hat{U}\hat{a}_{1,\text{in}}^\dagger\hat{U}^\dagger\right\} + \frac{\alpha_2}{\sqrt{2}}\left\{\hat{U}\hat{a}_{1,\text{in}}^\dagger\hat{U}^\dagger\right\}^2\right)\left\{\hat{U}\hat{a}_{2,\text{in}}^\dagger\hat{U}^\dagger\right\}\left|0,0,0\right\rangle_{\text{out}} \tag{40}$$

where we have added curly brackets to highlight the similarity transformations that carry the input operators to the output ones. We write these linear transformations as [1]

$$\hat{a}_{j,\text{in}}^\dagger \to \hat{U}\hat{a}_{j,\text{in}}^\dagger\hat{U}^\dagger = \sum_{k=1}^{3} S_{jk}^T \hat{a}_{k,\text{out}}^\dagger = \sum_{k=1}^{3} S_{kj} \hat{a}_{k,\text{out}}^\dagger \tag{41}$$

where the apparently strangely defined (hold on a while, the notation is rationalized below) coefficients $S_{jk}^T$ encode the reliance of the input→output operator transformation on the system parameters labeled in Fig. (2b). In terms of these coefficients, we can write after some algebra and after accounting for the matrix transpose by reordering the indices of matrix elements,

$$\left|\Psi'\right\rangle = \left[\alpha_0 S_{22}\left|0\right\rangle_1 + \alpha_1\left(S_{11}S_{22} + S_{21}S_{12}\right)\left|1\right\rangle_1 + \alpha_1 S_{11}\left(S_{11}S_{22} + 2S_{21}S_{12}\right)\left|2\right\rangle_1\right] \otimes \left|1\right\rangle_2 \otimes \left|0\right\rangle_3$$
$$+\left\{\alpha_0 \sum_{j\neq 2} S_{j2}\hat{a}_{j,\text{out}}^\dagger + \alpha_1 \sum_{(j,k)\neq\{(1,2),(2,1)\}} S_{j1}S_{k2}\hat{a}_{j,\text{out}}^\dagger\hat{a}_{k,\text{out}}^\dagger + \frac{\alpha_2}{\sqrt{2}} \sum_{(j,k,l)\neq\{\text{perm}(1,1,2)\}} S_{j1}S_{k1}S_{l2}\hat{a}_{j,\text{out}}^\dagger\hat{a}_{k,\text{out}}^\dagger\hat{a}_{l,\text{out}}^\dagger\right\}\left|0,0,0\right\rangle$$
(42)

where the branch of the state vector due to the part in curly brackets is ultimately rejected by the projective measurement with probability $p_{\text{fail}}^{NLPSG} = 1 - \left|\beta\right|^2$, see Eq. (32). Recalling the normalization condition on the coefficients $\alpha_j$ and comparing Eq. (42) with Eqs. (32) and (38), we arrive at the following constraints that determine the successful implementation of the NLPSG with $p_{\text{sucess}}^{NLPSG} = \left|\beta\right|^2$.

$$\alpha_0 S_{22} = \beta\alpha_0 \tag{43}$$

$$\alpha_1 \left( S_{11} S_{22} + S_{21} S_{12} \right) = \beta \alpha_1 \tag{44}$$

$$\alpha_2 S_{11} \left( S_{11} S_{22} + 2 S_{21} S_{12} \right) = -\beta \alpha_2. \tag{45}$$

### *Derivation of Transfer Matrices, $\mathbf{T}^{(j)}$ and Scattering Matrix $\mathbf{S}$*

In order to analyze the operation of the scalable NLPSG shown in Fig. (2a), we must work out the coefficients $S_{ij}$ in order to get $S_{ij}^T = S_{ji}$ as above, that describe the linear transformation of creation operators that characterizes the unitary evolution of the three-mode field through the device. Previously, we have proposed the directionally coupled double bus MRR as a fundamental circuit element for optical quantum information processing in silicon nanophotonics [2]. Comparing Fig. (2a) with Fig. (1) from Ref. [2], it is clear that our proposed NLPSG is an integrated network of three such fundamental circuit elements. To aid in the discussion we group the NLPSG into three "blocks," with each block involving a single MRR based circuit element coupling two of the modes; the remaining mode involves a linear phase shift across any given block. For example, regarding modes 2 and 3 of block 1 in Fig. (2a), the input Boson operators, $\hat{a}_{3,\text{in}}$ and $\hat{b}_{21}$, to the MRR are analogous to the operators $\hat{a}$ and $\hat{f}$, respectively, in Ref. [2], with a similar analogy for the outputs $\hat{a}_{12}, \hat{a}_{2,\text{out}} \leftrightarrow \hat{c}, \hat{l}$. Along mode 1, $\hat{c}_{12} = e^{-i\delta_1} \hat{a}_{1,\text{in}}$. Boson operators carrying two subscripts are internal to the NLPSG; we will eliminate them algebraically in deriving the operator input/output relations for the device.

Adapted to the notation we have adopted here the input/output operator transformations for the individual fundamental circuit elements implicated in Fig. (2a) can be written as

$$\begin{pmatrix} \hat{a}_{2,\text{out}} \\ \hat{a}_{12} \end{pmatrix} = \mathbf{M}^{(1)} \begin{pmatrix} \hat{b}_{21} \\ \hat{a}_{1,\text{in}} \end{pmatrix}, \quad \begin{pmatrix} \hat{c}_{23} \\ \hat{b}_{21} \end{pmatrix} = \mathbf{M}^{(2)} \begin{pmatrix} \hat{c}_{12} \\ \hat{b}_{32} \end{pmatrix}, \quad \begin{pmatrix} \hat{b}_{32} \\ \hat{a}_{3,\text{out}} \end{pmatrix} = \mathbf{M}^{(3)} \begin{pmatrix} \hat{a}_{2,\text{in}} \\ \hat{a}_{23} \end{pmatrix} \tag{46}$$

Where the superscript on the $2 \times 2$ matrix $\mathbf{M}^{(j)}$ labels the MRR with which it corresponds. Each of these matrices has the form

$$\mathbf{M}^{(j)} = \begin{pmatrix} A_j & B_j \\ C_j & D_j \end{pmatrix} \tag{47}$$

having matrix elements [2]

$$A_j = \frac{\eta_j - \tau_j^* e^{-i\theta_j}}{1 - \eta_j^* \tau_j^* e^{-i\theta_j}}, \qquad B_j = -\frac{\gamma_j \kappa_j^* e^{-i\phi_j}}{1 - \eta_j^* \tau_j^* e^{-i\theta_j}} \tag{48}$$

$$C_j = -\frac{\kappa_j \gamma_j^* e^{-i(\theta_j - \phi_j)}}{1 - \eta_j^* \tau_j^* e^{-i\theta_j}}, \qquad D_j = \frac{\tau_j - \eta_j^* e^{-i\theta_j}}{1 - \eta_j^* \tau_j^* e^{-i\theta_j}} \tag{49}$$

that depend explicitly on the system parameters, namely, the round trip phase shifts $(\theta_j)$, the direct transmission amplitudes $(\tau_j, \eta_j)$, and the cross coupling amplitudes $(\kappa_j, \gamma_j)$. The other phase shifts $\phi_j$ represents the phase partition induced by the specific locations of the couplings of the rings with the waveguides; the phase partitions have no effect on our results, so we implicitly set them to $\phi_j = \theta_j/2$ (symmetrically coupled rings).

We can now write the three mode input/output operator transformations for the three individual blocks of the NLPSG as

$$\begin{pmatrix} \hat{c}_{12} \\ \hat{a}_{2,\text{out}} \\ \hat{a}_{12} \end{pmatrix} = \mathbf{T}^{(1)} \begin{pmatrix} \hat{a}_{1,\text{in}} \\ \hat{b}_{21} \\ \hat{a}_{3,\text{in}} \end{pmatrix} \tag{50}$$

$$\begin{pmatrix} \hat{c}_{23} \\ \hat{b}_{21} \\ \hat{a}_{23} \end{pmatrix} = \mathbf{T}^{(2)} \begin{pmatrix} \hat{c}_{12} \\ \hat{b}_{32} \\ \hat{a}_{12} \end{pmatrix} \tag{51}$$

$$\begin{pmatrix} \hat{a}_{1,\text{out}} \\ \hat{b}_{32} \\ \hat{a}_{3,\text{out}} \end{pmatrix} = \mathbf{T}^{(3)} \begin{pmatrix} \hat{c}_{23} \\ \hat{a}_{2,\text{in}} \\ \hat{a}_{23} \end{pmatrix} \tag{52}$$

where the transfer matrices have block diagonal structure

$$\mathbf{T}^{(1,3)} = e^{i\delta_{1,3}} \mathbf{I}_1 \oplus \mathbf{M}^{(1,3)} \tag{53}$$

$$\mathbf{T}^{(2)} = \mathbf{M}^{(12)} \oplus e^{i\delta_2} \mathbf{I}_3. \tag{54}$$

The symbol $\mathbf{I}_k$ in Eqs. (53) and (54) represents the $1\times 1$ identity matrix appropriate to mode $k$.

It is clear that, according to Eqs.(50) through (52) that there is no direct algebraic substitution that will result in an operator input/output relation of the form we desire, namely,

$$\begin{pmatrix} \hat{a}_{1,\text{out}} \\ \hat{a}_{2,\text{out}} \\ \hat{a}_{3,\text{out}} \end{pmatrix} = \mathbf{S} \begin{pmatrix} \hat{a}_{1,\text{in}} \\ \hat{a}_{2,\text{in}} \\ \hat{a}_{3,\text{in}} \end{pmatrix} \tag{55}$$

where the matrix $\mathbf{S}$ describes the unitary "scattering" of the input operators into the output ones. Instead, owing to the directional nature of the couplings between waveguides and MRRs and to the topology of each MRR itself, we must algebraically adjust the relations encoded in the transfer matrices, $\mathbf{T}^{(j)}$, via Eqs.(50) through (52) by introducing a set of three mode swap operations as follows.

Let the matrix, $\mathbf{G}$, having elements $g_{ij}$, represent an arbitrary element of the group, $GL(3)$, of general linear transformations on three independent coordinates, $(x, y, z)$ such that $(x, y, z) \xrightarrow{GL(3)} (x', y', z')$ via

$$\begin{pmatrix} x' \\ y' \\ z' \end{pmatrix} = \mathbf{G} \begin{pmatrix} x \\ y \\ z \end{pmatrix} = \begin{pmatrix} g_{11} & g_{12} & g_{13} \\ g_{21} & g_{22} & g_{23} \\ g_{31} & g_{32} & g_{33} \end{pmatrix} \begin{pmatrix} x \\ y \\ z \end{pmatrix} = \begin{pmatrix} g_{11}x + g_{12}y + g_{13}z \\ g_{21}x + g_{22}y + g_{23}z \\ g_{31}x + g_{32}y + g_{33}z \end{pmatrix} \quad (56)$$

and define the three operations $\mathbb{S}_i^{(3)}[\mathbf{G}]$ such that

$$\mathbb{S}_1^{(3)}[G] \equiv \frac{1}{g_{11}} \begin{pmatrix} 1 & -g_{12} & -g_{13} \\ g_{21} & m_{3,3} & m_{3,2} \\ g_{31} & m_{2,3} & m_{2,2} \end{pmatrix} \quad (57)$$

$$\mathbb{S}_2^{(3)}[G] \equiv \frac{1}{g_{22}} \begin{pmatrix} m_{3,3} & g_{12} & -m_{3,1} \\ -g_{21} & 1 & -g_{23} \\ -m_{1,3} & g_{32} & m_{1,1} \end{pmatrix} \quad (58)$$

$$\mathbb{S}_3^{(3)}[G] \equiv \frac{1}{g_{33}} \begin{pmatrix} m_{2,2} & m_{2,1} & g_{13} \\ m_{1,2} & m_{1,1} & g_{23} \\ -g_{31} & -g_{32} & 1 \end{pmatrix} . \quad (59)$$

We have introduced in Eqs. (57)-(59) the standard minors, $m_{i,j}$, defined as the result of eliminating row $i$ and column $j$ from $\mathbf{G}$ and computing the determinant of the resulting $2 \times 2$ sub-matrix. With these operations, we can "swap" independent variables for dependent ones according to

$$\begin{pmatrix} x \\ y' \\ z' \end{pmatrix} = \mathbb{S}_1^{(3)}[\mathbf{G}] \begin{pmatrix} x' \\ y \\ z \end{pmatrix}, \quad \begin{pmatrix} x' \\ y \\ z' \end{pmatrix} = \mathbb{S}_2^{(3)}[\mathbf{G}] \begin{pmatrix} x \\ y' \\ z \end{pmatrix}, \text{ and } \begin{pmatrix} x' \\ y' \\ z' \end{pmatrix} = \mathbb{S}_2^{(3)}[\mathbf{G}] \begin{pmatrix} x \\ y \\ z' \end{pmatrix} \quad (60)$$

We will refer to the operations $\mathbb{S}_j^{(3)}[\mathbf{G}]$ as Mode Swap (MS) operations on the $j^{\text{th}}$ mode of the three mode input/output linear optical system. Specifically, consider MS operations on mode 2 for each of the blocks of the circuit in Fig. (2a). Appealing to Eqs. (50) through (52), (57), and (58), we can write for block 1

$$\begin{pmatrix} \hat{c}_{12} \\ \hat{a}_{2,\text{out}} \\ \hat{a}_{12} \end{pmatrix} = \mathbf{T}^{(1)} \begin{pmatrix} \hat{a}_{1,\text{in}} \\ \hat{b}_{21} \\ \hat{a}_{3,\text{in}} \end{pmatrix} \xrightarrow{\text{MS2}} \begin{pmatrix} \hat{c}_{12} \\ \hat{b}_{21} \\ \hat{a}_{12} \end{pmatrix} = \mathbb{S}_2^{(3)}\left[\mathbf{T}^{(1)}\right] \begin{pmatrix} \hat{a}_{1,\text{in}} \\ \hat{a}_{2,\text{out}} \\ \hat{a}_{3,\text{in}} \end{pmatrix}, \quad (61)$$

for block 2,

$$\begin{pmatrix} \hat{c}_{23} \\ \hat{b}_{21} \\ \hat{a}_{23} \end{pmatrix} = \mathbf{T}^{(2)} \begin{pmatrix} \hat{c}_{12} \\ \hat{b}_{32} \\ \hat{a}_{12} \end{pmatrix} \xrightarrow{MS2} \begin{pmatrix} \hat{c}_{23} \\ \hat{b}_{32} \\ \hat{a}_{23} \end{pmatrix} = \mathbb{S}_2^{(3)}\left[\mathbf{T}^{(2)}\right] \begin{pmatrix} \hat{c}_{12} \\ \hat{b}_{21} \\ \hat{a}_{12} \end{pmatrix}, \qquad (62)$$

and for block 3,

$$\begin{pmatrix} \hat{a}_{1,out} \\ \hat{b}_{32} \\ \hat{a}_{3,out} \end{pmatrix} = \mathbf{T}^{(3)} \begin{pmatrix} \hat{c}_{23} \\ \hat{a}_{2,in} \\ \hat{a}_{23} \end{pmatrix} \xrightarrow{MS2} \begin{pmatrix} \hat{a}_{1,out} \\ \hat{a}_{2,in} \\ \hat{a}_{3,out} \end{pmatrix} = \mathbb{S}_2^{(3)}\left[\mathbf{T}^{(3)}\right] \begin{pmatrix} \hat{c}_{23} \\ \hat{b}_{32} \\ \hat{a}_{23} \end{pmatrix}. \qquad (63)$$

We define the mode swap operation less for any reason motivated by physics and more for the satisfaction of human convenience, for upon staring at Eqs. (61) - (63), it becomes clear that we can now consistently substitute Eq. (61) into (62) and then the proceeds of that step into Eq. (63), resulting in

$$\begin{pmatrix} \hat{a}_{1,out} \\ \hat{a}_{2,in} \\ \hat{a}_{3,out} \end{pmatrix} = \mathbb{S}_2^{(3)}\left[\mathbf{T}^{(3)}\right] \mathbb{S}_2^{(3)}\left[\mathbf{T}^{(2)}\right] \mathbb{S}_2^{(3)}\left[\mathbf{T}^{(1)}\right] \begin{pmatrix} \hat{a}_{1,in} \\ \hat{a}_{2,out} \\ \hat{a}_{3,in} \end{pmatrix}. \qquad (64)$$

Another MS 2 operation produces

$$\begin{pmatrix} \hat{a}_{1,out} \\ \hat{a}_{2,out} \\ \hat{a}_{3,out} \end{pmatrix} = \mathbb{S}_2^{(3)}\left[ \mathbb{S}_2^{(3)}\left[\mathbf{T}^{(3)}\right] \mathbb{S}_2^{(3)}\left[\mathbf{T}^{(2)}\right] \mathbb{S}_2^{(3)}\left[\mathbf{T}^{(1)}\right] \right] \begin{pmatrix} \hat{a}_{1,in} \\ \hat{a}_{2,in} \\ \hat{a}_{3,in} \end{pmatrix}, \qquad (65)$$

allowing us to identify the $\mathbf{S}$ matrix encoding the operator transformations for the global linear optical network in terms of transfer matrices, $\mathbf{T}^{(j)}$, describing the local input/output relations for each of the individual blocks of the circuit, specifically,

$$\mathbf{S} = \mathbb{S}_2^{(3)}\left[ \mathbb{S}_2^{(3)}\left[\mathbf{T}^{(3)}\right] \mathbb{S}_2^{(3)}\left[\mathbf{T}^{(2)}\right] \mathbb{S}_2^{(3)}\left[\mathbf{T}^{(1)}\right] \right]. \qquad (66)$$

The mode swap operation we have introduced in the foregoing calculation can be generalized for application to an $N$ mode linear optical system comprised of directionally coupled waveguides and MRRs in a very straightforward fashion. One can work out using $M^{th}$ grade algebra, where, hopefully, $M \leq 8$, the form $\mathbb{S}_j^{(N)}[\mathbf{G}]$ on the linear transformation $GL(N)$ that accomplishes the MS on mode $j$ for $j \leq N$. The results for any $N$ can be expressed in a reasonably compact form by introducing the higher order minors $m_{\underset{N-2\,\text{rows}}{klm...},\underset{N-2\,\text{cols}}{pqr...}}$ of $\mathbf{G}$. We will present elsewhere this procedure, its underlying multi-linear algebraic structure, and examples of its use.

Working out the unitary part of the evolution using the Heisenberg Picture [3], we actually want to express the input creation operators in terms of the output ones,

$$\begin{pmatrix} \hat{a}^{\dagger}_{1,\text{in}} \\ \hat{a}^{\dagger}_{2,\text{in}} \\ \hat{a}^{\dagger}_{3,\text{in}} \end{pmatrix} = \left(\mathbf{S}^{-1}\right)^{*} \begin{pmatrix} \hat{a}^{\dagger}_{1,\text{out}} \\ \hat{a}^{\dagger}_{2,\text{out}} \\ \hat{a}^{\dagger}_{3,\text{out}} \end{pmatrix}. \qquad (67)$$

The Bosonic commutation relations $\left[\hat{a}_{j,\text{out}}, \hat{a}^{\dagger}_{k,\text{out}}\right] = \delta_{jk}$, $\left[\hat{a}_{j,\text{out}}, \hat{a}_{k,\text{out}}\right] = \left[\hat{a}^{\dagger}_{j,\text{out}}, \hat{a}^{\dagger}_{k,\text{out}}\right] = 0$, with similar relations for the input operators, constrain the $\mathbf{S}$ matrix to be unitary, $\mathbf{S}^{-1} = \mathbf{S}^{\dagger} = \left(\mathbf{S}^{T}\right)^{*}$, which, in turn, implies that $\left(\mathbf{S}^{-1}\right)^{*} = \mathbf{S}^{T}$ so that

$$\begin{pmatrix} \hat{a}^{\dagger}_{1,\text{in}} \\ \hat{a}^{\dagger}_{2,\text{in}} \\ \hat{a}^{\dagger}_{3,\text{in}} \end{pmatrix} = \mathbf{S}^{T} \begin{pmatrix} \hat{a}^{\dagger}_{1,\text{out}} \\ \hat{a}^{\dagger}_{2,\text{out}} \\ \hat{a}^{\dagger}_{3,\text{out}} \end{pmatrix} \qquad (68)$$

as in Eq. . (41)

### *The role of non-trivial inter-MRR phase shift $\delta_2 \neq 0 \bmod 2\pi$*

Here we explore the role of the non-trivial inter-MRR phase shift $\delta_2 \neq 0 \bmod 2\pi$ while keeping all the MRRs on-resonance ($\theta_i = 0 \bmod 2\pi$) and again using $t_i = \mathcal{T}_i$ to ensure that $|\beta|^2 = 1/4$. The analysis is much more involved now since $A_2 = |A_2|\exp[i\theta_{A_2}]$ is complex with non-zero phase $\theta_{A_2}$. The solution is developed in [4] and involves the intersection of the 2D surface of the form of Eq.(58) (now with index $i$=2) with the 2D surface $\theta_{A_2}(\eta_2, \tau_2, \theta_2) = -\delta_2$ for a given value of $\delta_2 \neq 0$. These two surfaces intersect on a 1D manifold that can be numerically found, and representatively shown in Fig 5 for $\delta_2 = \pi/30$ (near-balanced inter-MRR waveguide phase). To avoid the reduction to this lower 1D manifold (with a reduced number of solutions over that of Fig. 3 (in the main text) with $i$=2), one would want to adjust $\delta_2 \to 0$, which could be achieved operationally by applying, say thermal heating to electrodes placed over this inter-MRR waveguide.

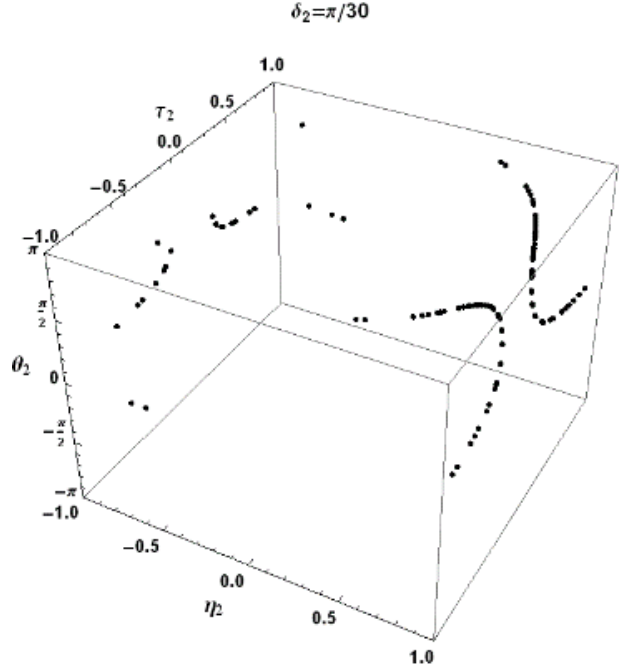

**Figure S1**: The 1D manifold resulting from the intersection of two 2D-manifolds on which optimal operation ($|\beta|^2 = 1/4$) of the scalable NLPSG occurs under conditions of on-resonance $\theta_i$=0mod2π for all MRRs, $\delta_{1,3}$=0mod2π for the outer waveguide phases, but with $\delta_2$=π/30mod2π phase shifts for the inter-MRR waveguide.

## References

[1]  J. Skaar, J. C. G. Escartín, and H. Landro, *Am. J. Phys*. **72**, 1385 (2004).
[2]  E. E. Hach III, S. F. Preble, A.W. Elshaari, P. M. Alsing, and M. L. Fanto, *Phys. Rev*. **A89**, 043805 (2014).
[3]  C. C. Gerry and P. L. Knight, *Introductory Quantum Optics* (Cambridge, 2005).
[4]  P. M. Alsing and E. E. Hach III, DOI: 10.1117/12.2325080, Quantum Technologies and Quantum Information Science **IV**, Berlin (Sep, 2018).